\documentclass[journal]{IEEEtran}
\usepackage{subfiles}
\usepackage{graphicx}
\usepackage{amsmath,amssymb,amsfonts}
\usepackage{color,colortbl}

\usepackage{acronym}
\acrodef{VT}{ventricular tachycardia}
\acrodef{VF}{ventricular fibrillation}
\acrodef{SVT}{supraventricular tachycardia}
\acrodef{AF}{atrial fibrillation}
\acrodef{PRLW}[PRL+W]{PR Logic + Wavelet}
\acrodef{EGM}{electrogram}
\acrodef{ECG}[ECG]{electrocardiogram}
\acrodef{ICD}[ICD]{implantable cardioverter defibrillator}
\acrodef{AGC}[AGC]{automatic gain control}
\acrodef{AAS}[AAS]{auto-adjusting sensitivity}
\acrodef{NSR}[NSR]{normal sinus rhythm}
\acrodef{FPR}[FPR]{false positive rate}
\acrodef{RIGHT}{Rhythm ID Goes Head-to-head Trial}
\acrodef{FDA}{U.S. Food and Drug Administration}
\acrodef{CPS}{cyber-physical systems}
\acrodef{MRI}{magnetic resonance imaging}
\acrodef{BSC}{Boston Scientific}
\acrodef{MDT}{Medtronic}
\acrodef{VCL}{ventricular cycle length}
\acrodef{MCPS}{medical cyber-physical system }
\acrodef{V2}{Vitality 2}
\acrodef{BBN}{Bayesian belief network}
\acrodef{CACT-RIGHT}{Computer-Aided Clinical Trial for RIGHT}
\acrodef{GSN}{goal-structuring notation}
\acrodef{SR}{sinus rhythm}
\acrodef{EAM}{electroanatomic mapping}
\acrodef{ROI}{region of interest}
\acrodef{GP}{Gaussian processes}
\acrodef{GPR}{Gaussian process regression}
\acrodef{PV}{pulmonary vein}
\acrodef{LVA}{low voltage area}
\acrodef{AUC}{area under the curve}
\acrodef{LA}{left atrial}
\acrodef{CA}{catheter ablation}
\graphicspath{{figure/}}

\begin{document}

\title{Electroanatomic Mapping to determine Scar Regions in patients with Atrial Fibrillation}

\author{Jiyue He,$^{1*}$ Kuk Jin Jang,$^{1*}$  Katie Walsh, M.D.,$^{2}$  Jackson Liang, M.D.,$^{2}$ Sanjay Dixit, M.D.,$^{2}$ Rahul Mangharam$^{1}$
\thanks{$^{1}$ University of Pennsylvania, School of Engineering and Applied Science, Department of Electrical and Systems Engineering. $^{2}$ Hospital of the University of Pennsylvania, Department of Cardiac Electrophysiology. $^{*}$ Both authors contributed equally to this work.}
\thanks{July 23, 2019}}

\maketitle

\begin{abstract}
Left atrial voltage maps are routinely acquired during electroanatomic mapping  in patients undergoing catheter ablation for atrial fibrillation. For patients, who have prior catheter ablation when they are in sinus rhythm, the voltage map can be used to identify low voltage areas using a threshold of 0.2 - 0.45 mV. However, such a voltage threshold for maps acquired during atrial fibrillation has not been well established. A prerequisite for defining a voltage threshold is to maximize the topologically matched low voltage areas between the electroanatomic mapping acquired during atrial fibrillation and sinus rhythm. This paper demonstrates a new technique to improve the sensitivity and specificity of the matched low voltage areas. This is achieved by computing omni-directional bipolar voltages and applying Gaussian Process Regression based interpolation to derive the AF map. The proposed method is evaluated on a test cohort of 7 male patients, and a total of 46,589 data points were included in analysis. The low voltage areas in the posterior left atrium and pulmonary vein junction are determined using the standard method and the proposed method. Overall, the proposed method showed patient-specific sensitivity and specificity in matching low voltage areas of 75.70\% and 65.55\% for a geometric mean of 70.69\%. On average, there was an improvement of 3.00\% in the geometric mean, 7.88\% improvement in sensitivity, 0.30\% improvement in specificity compared to the standard method. The results show that the proposed method is an improvement in matching low voltage areas. This may help develop the voltage threshold to better identify low voltage areas in the left atrium for patients in atrial fibrillation.
\end{abstract}

\section{Introduction}
The voltage map is one form of data provided by \acf{EAM} and is often used to demarcate \acfp{LVA} and preserved voltage areas during catheter ablation therapy to treat \acf{AF}. \acp{LVA} correspond to areas of diseased atrium (fibrosis) or dense scars from prior ablations. Identifying \ac{LVA} can help in  planning ablation strategies especially in patients requiring repeat ablation procedures for arrhythmia recurrences, for example, \ac{AF} and atypical atrial flutter.

The cutoff threshold voltage for a voltage map that determines \acp{LVA} has been established for maps obtained while the patient is in \acf{SR} \cite{Kapa_2014}\cite{Squara_2014}. However, this same threshold is not applicable to maps collected during \ac{AF}.

\begin{figure}[ht]
	\centering
	\includegraphics[width=0.5\textwidth]{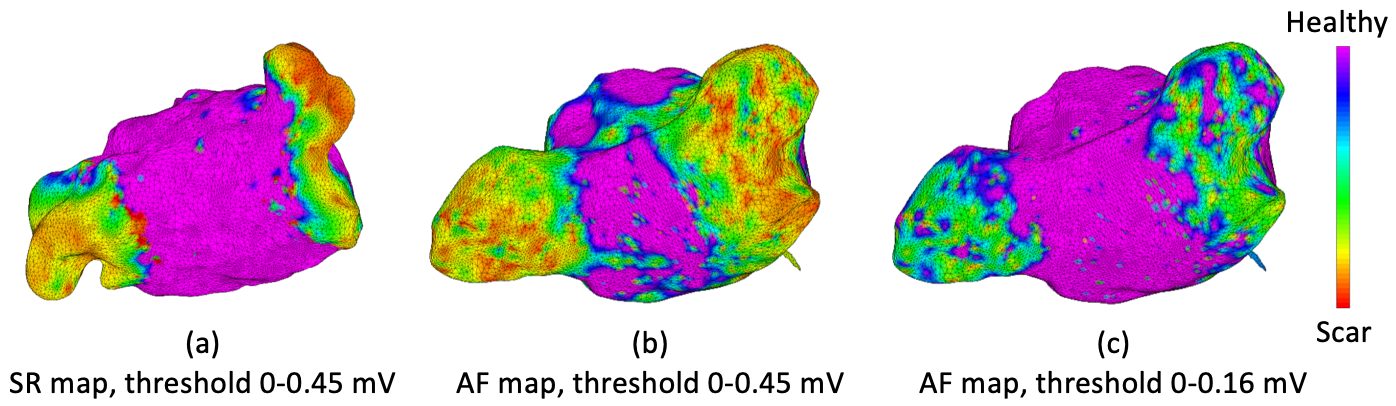}
	\caption{Example of need for a threshold to be used on voltage maps obtained during \ac{AF}. Magenta area is healthy tissue, other colors are scar tissue. (a) The \ac{SR} cutoff threshold of 0.45 mV is applied to the SR map. (b) Applying the same threshold to the AF map distorts \acp{LVA} in the AF map. (c) Adjusting the threshold to 0.16 mV restores the \acp{LVA}.}
	\label{fig:threshold_problem}
\end{figure}

For example, Fig. \ref{fig:threshold_problem}(a) depicts the \ac{SR} map with a cutoff threshold of 0.45 mV. The regions below the threshold are clearly delineated from the healthy regions (magenta). However, in (b), applying the same threshold to the \ac{AF} map for the same patient distorts the \acp{LVA} distribution. After adjusting to a lower threshold, as in (c), the \acp{LVA} on the \ac{AF} map are restored to match the \ac{SR} map.

Several studies have shown that the local atrial signal acquired during \ac{AF} is lower than in \ac{SR}. Thus, identifying \ac{LVA} during \ac{EAM} in \ac{AF} should require a lower cutoff voltage \cite{Rodriguez_2018}. However, determining a consistent threshold that can be applied to all patients remains challenging. A prerequisite is determining the best match of \acp{LVA} that can be obtained between the SR map and the AF map and thereby finding the best threshold to be applied on a patient-by-patient basis.

Problem Statement: Given a set of measurements during \ac{SR} and \ac{AF} for a patient, maximize the topologically matched \acp{LVA} between the derived \ac{SR} and \ac{AF} map and determine the best patient-specific cutoff voltage threshold.

In this paper, we demonstrate a method of deriving the \ac{AF} map which is robust to noise and error in the measurements and improves the patient-specific sensitivity and specificity of matched \acp{LVA} in comparison to the standard method through the following the contributions: 

\noindent\begin{itemize}
\item 
		Compute omni-directional bipolar voltages which are invariant to the orientation of the catheter, thus improving signal strength during \ac{AF}.
\item 
		Apply \ac{GPR} interpolation which improves the accuracy of \ac{LVA} detection in regions of the atrium with lower measurement density. 
\end{itemize}

\section{Background: Standard voltage map}
\begin{figure}[ht]
	\centering
	\includegraphics[width=0.5\textwidth]{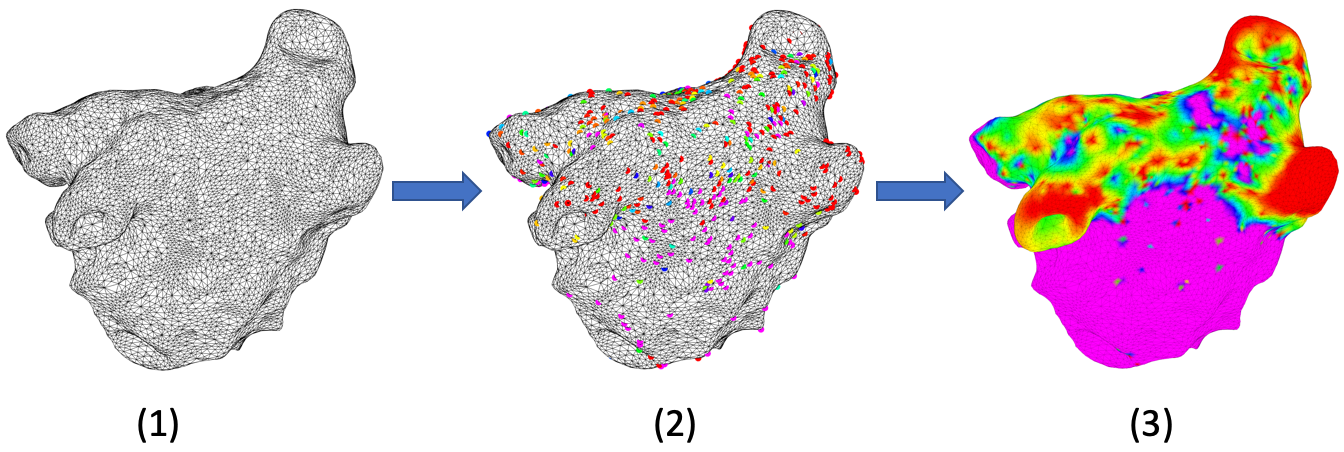}
	\caption{Standard method of deriving voltage map. (1) Anatomical mesh is computed from catheter locations. (2) Bipolar voltages are computed from measurements of electrogram at points along the mesh. (3) The bipolar voltages are interpolated to the remainder of the mesh to derive the voltage map.}
	\label{fig:standard_method}
\end{figure}

Fig. \ref{fig:standard_method} depicts the steps of deriving the current standard \ac{LA} voltage map during catheter ablation of \ac{AF}. Initially, as in (1), a 3D anatomical mesh is generated by manipulating a multi-electrode mapping catheter (Lasso or Pentaray) to different parts of the \ac{LA} \cite{Liang_2017}. As the mesh is being created, recordings of 2.5 seconds of electrogram are collected at various locations around the endocardium. (2) shows the locations of all such sampled points on the \ac{LA} mesh. The voltage value is interpolated to the remaining areas of the mesh to derive the final voltage map as shown in (3). The colors of the voltage map are based on a pre-specified cutoff threshold, where areas above the threshold are marked in magenta, and areas below the threshold are considered \acp{LVA}. 

\begin{figure}[ht]
	\centering
	\includegraphics[width=0.5\textwidth]{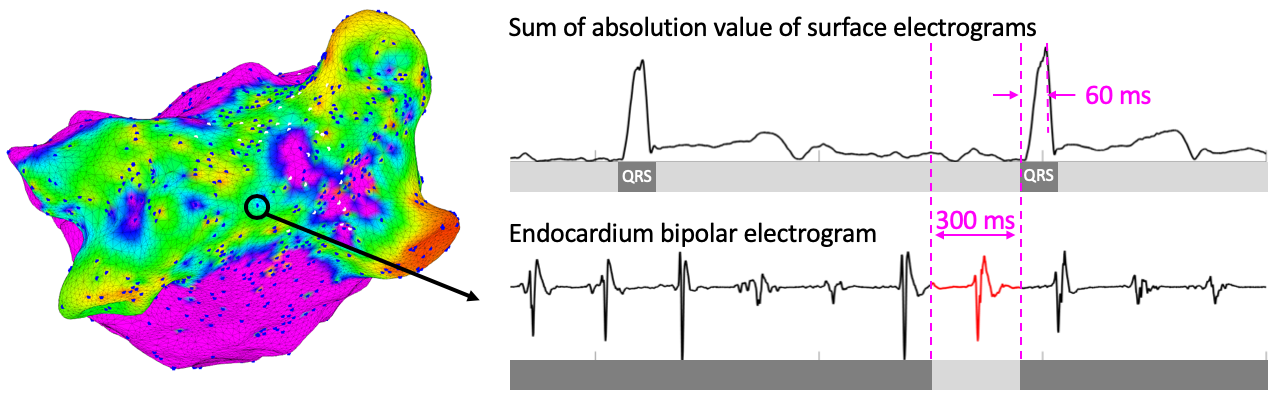}
	\caption{Bipolar voltages are computed as the peak-to-peak voltage in a pre-specified window within a 2.5 seconds recording of the endocardium electrogram.}
	\label{fig:electrode_sample_point}
\end{figure}

Fig. \ref{fig:electrode_sample_point} shows that within the 2.5 seconds, a 300 ms time window is defined relative to the QRS peak, and the peak-to-peak voltage is computed within the time window.

\section{Robust Method for Voltage Map Derivation}
\begin{figure}[ht]
	\centering
	\includegraphics[width=0.5\textwidth]{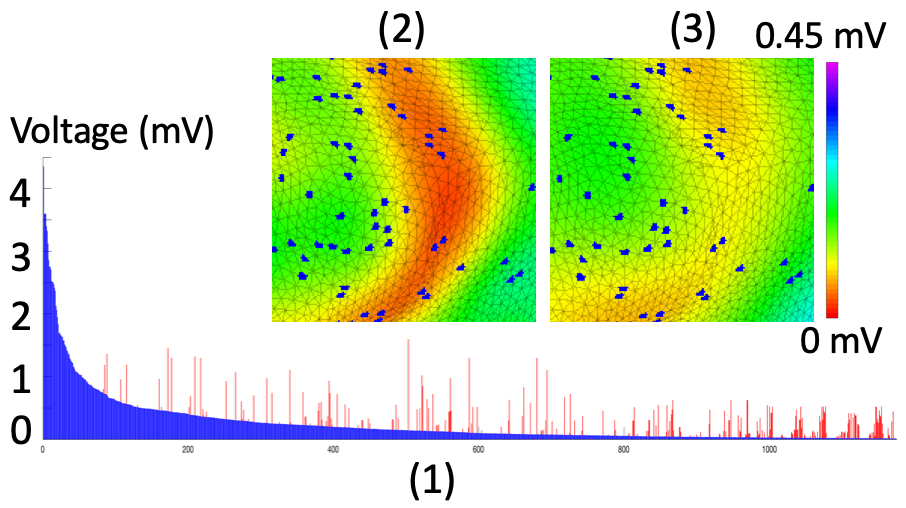}
	\caption{Benefits of using omni-directional bipolar voltages. (1) Bipolar voltages in \ac{AF} (blue) are amplified (red), improving the signal. (2), (3) Regions of previously low voltage (red in (2)) are increased after computing omni-directional bipolar voltages (yellow in (3)).}
	\label{fig:omni_overview}
\end{figure}

Our proposed robust method deriving the voltage map differs from the standard method in terms of two components: a) omni-directional bipolar voltage and b) \ac{GPR}-based interpolation.

a) Omni-directional bipolar voltage: Fig. \ref{fig:omni_overview} shows a  limitation of bipolar recording, the dependency on electrode orientation. If the bipolar placement is parallel to the iso-electric potential line, the bipolar recording will be zero, which does not reflect the local electric activity \cite {Magtibay_2017}. To reduce such dependency, we derive the omni-directional bipolar voltage. For each sample point, we select the unipolar electrogram recorded in the vicinity of the sample and compute all possible bipolar electrogram from this set. We approximate the omni-directional bipolar voltage as the largest bipolar amplitude from this set. Fig. \ref{fig:omni_overview} (1) Blue depicts the histogram of the voltages for a patient. Red shows how the corresponding voltages are amplified. In the tail portion, the voltages of some \ac{LVA} have increased above the threshold to be classified as healthy tissue. (2) and (3) exemplifies how the original red areas of voltage $\sim$0.06 mV are enhanced into yellow and green areas of voltage $\sim$0.17 mV. 

\begin{figure}[ht]
	\centering
	\includegraphics[width=0.5\textwidth]{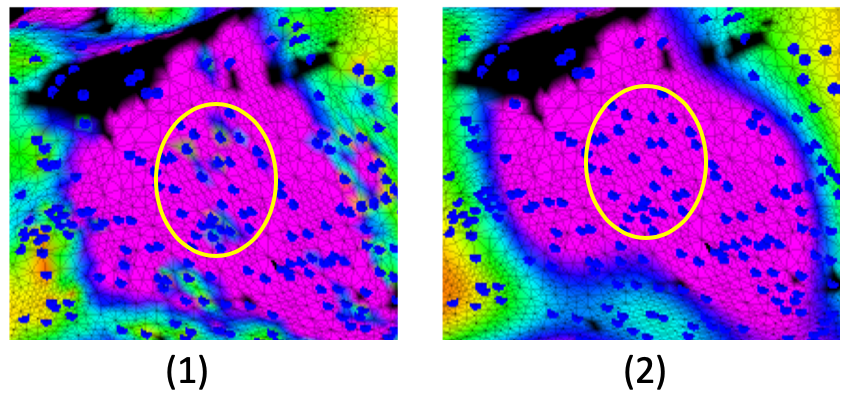}	
	\caption{Standard interpolation vs. \ac{GPR}-based interpolation. (1) Due to interpolation error, the circled region is determined as a \ac{LVA}. (2) \ac{GPR}-based interpolation accounts for measurements in the vicinity and reduces the \ac{LVA} in the circled region.}
	\label{fig:interpolation_limitation}
\end{figure}

\begin{figure*}[ht]
	\centering
	\includegraphics[width=1\textwidth]{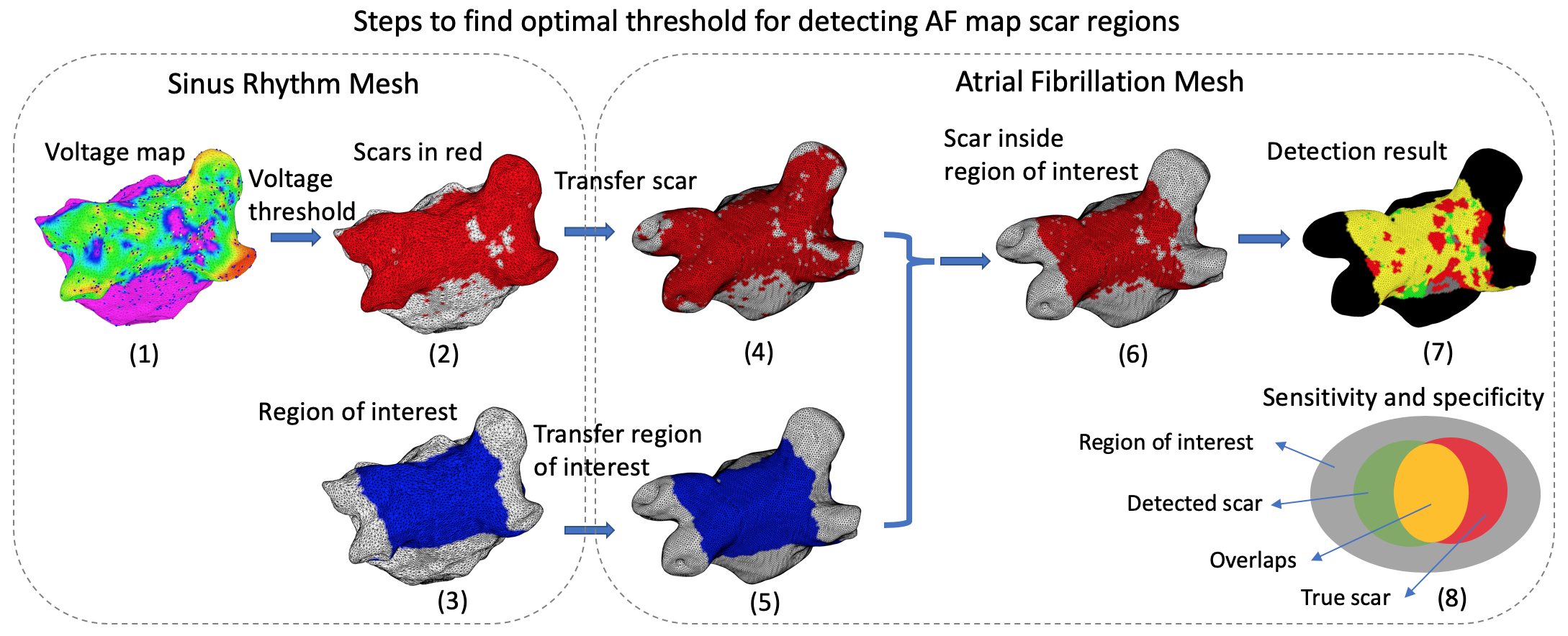}
	\caption{Experimental setup.  For each patient: From the SR map (1), \ac{LVA} are determined (2) as well as the region of interest (3). Both (2) and (3) are topologically transferred to the AF map (4) and (5). (4) (5) are intersected to obtain the true \acp{LVA} on the AF map (6). The optimal patient-specific threshold is determined by maximizing the product of sensitivity and specificity accordin to (8).}
	\label{fig:evaluation_overview}
\end{figure*}

b) Gaussian process regression based interpolation: If we model the endocardium as a surface and define the entire set of samples as, $\mathcal{D} = \{\mathbf{x}_n, y_n\}^N_{n=1}$, where inputs $\mathbf{X} = \{\mathbf{x}_n\}^N_{n=1}$ correspond to the locations on the mesh and  $\mathbf{y} = \{y_n\}^N_{n=1}$ are the voltage value at that location. Interpolating from these measured samples to the remainder of the mesh can be thought of as determining the estimates of the voltages at locations $\mathbf{X^*}$. The two major sources of interpolation error are the low measurement density and measurement noise. Both of these can be accounted for by modeling them using a Gaussian process $GP(m(\mathbf{x},k(\mathbf{x},\mathbf{x}')))$ \cite{Rasmussen_2006}, which is characterized by the mean $m(\mathbf{x})$ and the covariance $k(\mathbf{x},\mathbf{x}')$ kernel functions. We assume the common zero mean function and use the squared exponential function $k(\mathbf{x},\mathbf{x}')= \exp(-|| \mathbf{x}-\mathbf{x}' || / (2 \cdot l^2))$. Fig. \ref{fig:interpolation_limitation}(1) shows how the standard interpolation can result in regions that are classified as \acp{LVA}, due to interpolation error. (2) shows \ac{GPR}-based interpolation can improve the boundaries of \acp{LVA} by considering the surrounding measurements.

For determining the optimum threshold, a search in the range of 0-0.45 mV is performed to maximize the product of sensitivity and specificity. Here, sensitivity $= \frac{\text{TP}}{TP+FN}$ and specificity $= \frac{\text{TN}}{TN+FP}$, where a true positive (TP) indicates that the corresponding face on the anatomical mesh is detected as a \ac{LVA} and the true label is a \ac{LVA}.

The proposed method of voltage map derivation was evaluated on a test cohort of 7 patients that underwent repeat catheter ablation for \ac{AF} at the Hospital of the University of Pennsylvania. Patient demographics are: age $66\pm 0.7$ years, height $5'10" \pm 4"$, weight $238.9 \pm 50$ lb and ejection fraction $49.4\%$. Details are in Table \ref{table:patient_population}. For each patient, voltage maps were sequentially obtained using Carto3 (Biosense Webster) during \ac{SR} and \ac{AF} or vice-versa. Fill threshold was 5 mm, and filters were set at 2 to 240 Hz for unipolar \acp{EGM}, 16-500 Hz for bipolar \acp{EGM}, and 0.5-200 Hz for surface electrogram recordings. We  assumed that the LVA observed based on the \ac{SR} voltage map is the ground truth. 

\begin{table}[htb]
	\centering
	\begin{tabular}{ c|ccccccc } 
		\hline
		Patient & 1 & 2 & 3 & 4 & 5 & 6 & 7 \\ \hline
		Age & 66 & 58 & 70 & 70 & 69 & 58 & 76 \\
		Height (inch) & 71 & 75 & 72 & 62 & 70 & 72 & 68 \\
		Weight (pound) & 290 & 260 & 266 & 243  & 267 & 200 & 146 \\
		Ejection Function (\%) & 65 & 50 & 25 & 50 & 55 & 60 & 40 \\
		\hline
	\end{tabular}
	\vspace{2pt}
	\caption{Patient population}
	\label{table:patient_population}	
\end{table}

\section{Experimental setup}
The overall evaluation process is depicted in Fig. \ref{fig:evaluation_overview}. From the original data (1), we apply the standard 0.45 mV cutoff threshold on the SR map to identify regions of \ac{LVA} (2). We select a \ac{ROI} on the SR map (3), which consists of the posterior \ac{LA} and \ac{PV} junctions. Both the \ac{LVA} and the \ac{ROI} is transferred to the AF mesh and intersected to form the final \ac{LVA} on the AF map (4),(5),(6). Then result (7) is determined as shown in (8) in terms of sensitivity and specificity. 

\section{Results and discussion}
\begin{figure}[htb]
	\centering
	\includegraphics[width=0.5\textwidth]{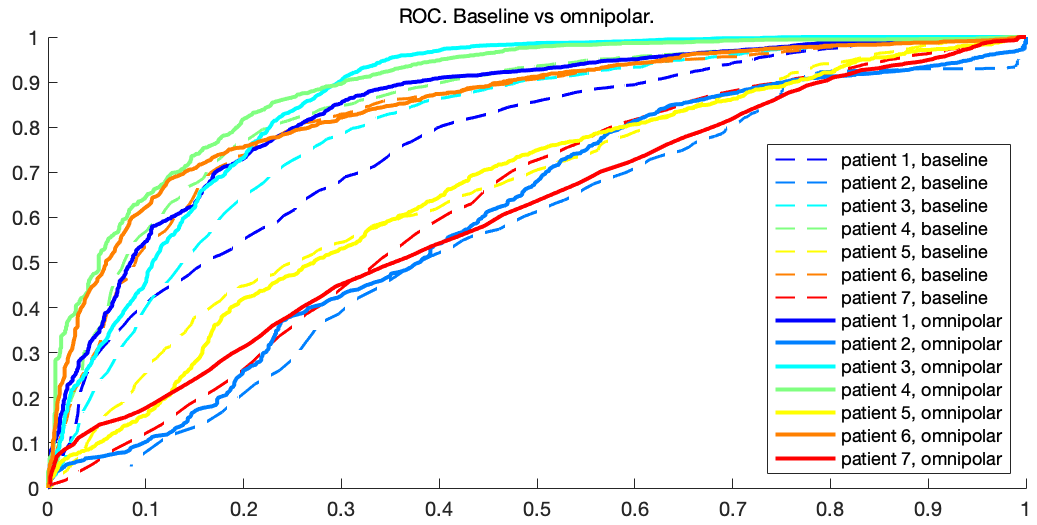}
	\caption{ROC curve comparing, baseline method vs proposed method using omni-directional bipolar voltages and \ac{GPR}-based interpolation. The proposed method shows improved or similar performance across various thresholds.}
	\label{fig:ROC_baseline_gaussian_omni_all_patients}
\end{figure}

A total of 46,589 data points were included in analysis, that was on average 6,656 data points for each of the 7 patients. Table \ref{table:patient_specific_result} summarizes the results of evaluation for each patient. On average, our proposed method showed a sensitivity and specificity of 75.70\% and 66.55\%, respectively. This was a 3.00\% improvement in the geometric mean compared to the standard method. Moreover, our proposed method exhibited a 7.88\% improvement in sensitivity and a 0.30\% improvement in specificity. ROC curves were obtained for each of the methods and the \acl{AUC} was computed as shown in Fig. \ref{fig:ROC_baseline_gaussian_omni_all_patients}. Our proposed method showed an average of 3.91\% improvement in terms of the \ac{AUC}. 

\definecolor{Gray}{gray}{0.85}

\begin{table*}[t]
	\centering
	\resizebox{1\textwidth}{!}{
		\begin{tabular}{|c|c|c|c|c|c|c|c|c|c|c|c|c|c|c|c|c|c} \hline
		\multicolumn{1}{|c|}{Patient} &\multicolumn{5}{c|}{Baseline} &\multicolumn{5}{c|}{Omni.+GP} &\multicolumn{4}{c|}{Percentage of Improvement}\\ \hline
		&  Sens. & Spec. & GM & AUC & Voltage Threshold &Sens. & Spec. & GM. & AUC & Voltage Threshold & $\Delta$Sens. & $\Delta$Spec. & $\Delta$GM & $\Delta$AUC\\ \hline
1 & 76.18 & 63.56 & 69.58 & 0.77  & 0.26  & 84.92 & 70.74 & 77.51 & 0.85  & 0.23  & 11.47 & 11.30 & 11.40 & 10.39 \\
2 & 54.87 & 57.64 & 56.24 & 0.57  & 0.09  & 72.72 & 47.49 & 58.77 & 0.61  & 0.11  & 32.53 & -17.61 & 4.50  & 7.02 \\
3 & 77.38 & 71.48 & 74.37 & 0.80  & 0.16  & 91.22 & 69.59 & 79.67 & 0.86  & 0.15  & 17.89 & -2.64 & 7.13  & 7.50 \\
4 & 79.52 & 78.08 & 78.80 & 0.85  & 0.26  & 85.42 & 77.07 & 81.14 & 0.89  & 0.30  & 7.42  & -1.29 & 2.97  & 4.71 \\
5 & 58.08 & 66.88 & 62.32 & 0.67  & 0.36  & 69.24 & 56.71 & 62.66 & 0.66  & 0.38  & 19.21 & -15.21 & 0.55  & -1.49 \\
6 & 80.61 & 74.64 & 77.57 & 0.83  & 0.32  & 72.87 & 83.41 & 77.96 & 0.85  & 0.21  & -9.60 & 11.75 & 0.50  & 2.41 \\
7 & 70.24 & 52.57 & 60.77 & 0.63  & 0.10  & 53.54 & 60.87 & 57.09 & 0.61  & 0.17  & -23.78 & 15.79 & -6.06 & -3.17 \\ \hline
\rowcolor{Gray} Average & 70.98 & 66.41 & 68.52 & 0.73  & 0.22  & 75.70 & 66.55 & 70.69 & 0.76  & 0.22  & 7.88  & 0.30  & 3.00  & 3.91 \\
Std  & 10.49 & 9.19  & 8.87  & 0.11  & 0.11  & 12.72 & 12.34 & 10.65 & 0.13  & 0.09  & 18.96 & 13.32 & 5.54  & 4.95 \\ \hline
		\end{tabular}
	}
	\vspace{2pt}
	\caption{Results for patient-specific performance for all patients}
	\label{table:patient_specific_result}
\end{table*}

\begin{figure}[ht]
	\centering
	\includegraphics[width=0.5\textwidth]{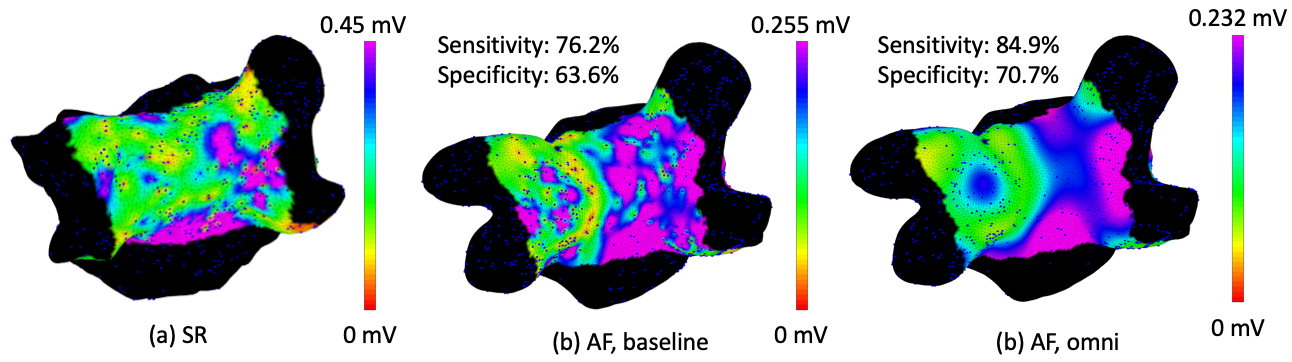}
	\caption{Example of improved result: Patient 1, baseline vs proposed method.}
	\label{fig:patient_1_baseline_omni}
\end{figure}

Fig. \ref{fig:patient_1_baseline_omni} depicts patient 1 with the improved detection of \acp{LVA} after applying the optimal threshold. First, the improvement is due to the enhanced signal in the omni-directional bipolar voltages. On the boundaries of the \ac{ROI} \acp{LVA} are correctly categorized as healthy regions using the proposed method. Second, in the areas of high electrode density, \ac{GPR}-based interpolation discounts the amplitude when voltage spikes due to noise occur. Finally, in regions of lower measurement density, the standard method underestimates the interpolated voltage when spurious low-voltage measurement exist. In this case the combination of omni-directional bipolar voltages enhances the signal, and \ac{GPR}-based interpolation filters such noise, preventing classification as \acp{LVA}. 

\begin{figure}[ht]
	\centering
	\includegraphics[width=0.5\textwidth]{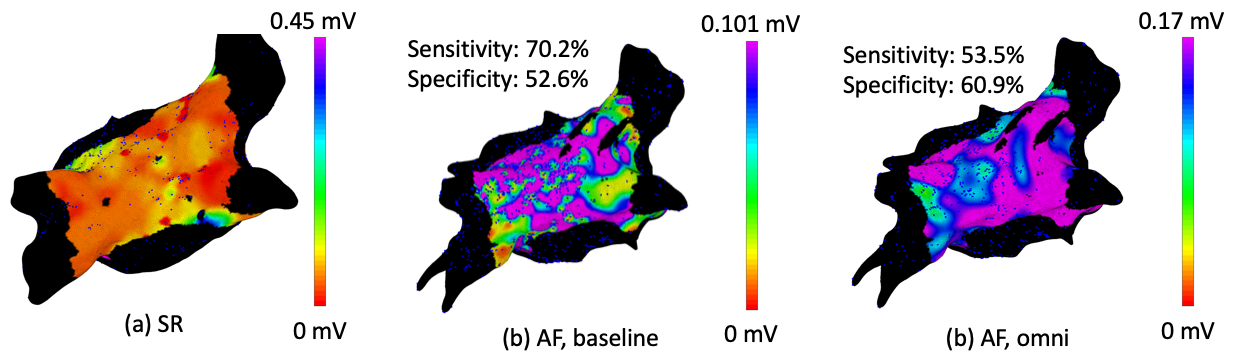}
	\caption{Patient 7. Baseline vs proposed method. Poor performance can be attributed to the dominance of \acp{LVA} in the \ac{ROI}, thus penalizing specificity.}
	\label{fig:patient_7_baseline_omni}
\end{figure}

In patient 7, our proposed method did not improve performance. Upon further inspection, we discovered that 90\% of the ROI was \ac{LVA} as shown in Fig. \ref{fig:patient_7_baseline_omni}. This patient had undergone prior extensive surgical ablation and so had extensive areas of dense scar in the \ac{ROI}, making discrimination difficult. Optimizing the threshold with a different criterion which accounts for this bias may result in a better outcome.

Limitations: The results of our study are limited in scope, mostly in part by the small cohort size. Bias due to the particular demographics of the cohort may have affected the results and different results may be obtained with a larger cohort. Selection of different hyperparameters for the method, such as a different kernel for \ac{GPR}-based regression may affect the performance. Finally, establishing a standard protocol for obtaining data may lead to improvement.

\section{Conclusion}
In this work, we have presented a method for deriving the voltage map during \ac{AF} and comparing it to voltage maps acquired during \ac{SR}. Our method computes omni-directional bipolar voltages from the measurements and utilizes \ac{GPR}-based interpolation to derive the voltage map. Evaluation on the test cohort showed that, in general, the method improved the patient-specific sensitivity and specificity in determining \acp{LVA} of the AF map compared to the standard method, though some exceptions exist. This improvement in matched areas between the maps is significant and has important practical implications as clinicians interpret voltage maps according to the areas and not by the individual point measurements. More accurate information about \ac{LVA} distribution is helpful to clinicians in planning ablation strategies for patients who require repeat catheter ablation for arrhythmia recurrences. Immediate future work is to apply the method over a larger cohort. In a practical clinical setting, patient-to-patient variability may need to be accounted for in the criterion. Overall, the results provide evidence that the proposed method improves the detection of \acp{LVA} in \ac{AF} maps. Because of the robustness to measurement noise and interpolation error, the proposed method could lead to a more consistent criterion.

\bibliographystyle{ieeetr}
\bibliography{reference}

\begin{thebibliography}{1}

\bibitem{Kapa_2014}
S.~Kapa {\em et~al.}, ``Contact electroanatomic mapping derived voltage
  criteria for characterizing left atrial scar in patients undergoing ablation
  for atrial fibrillation,'' {\em Journal of Cardiovascular Electrophysiology},
  vol.~25, no.~10, pp.~1044--1052, 2014.

\bibitem{Squara_2014}
F.~Squara {\em et~al.}, ``Voltage mapping for delineating inexcitable dense
  scar in patients undergoing atrial fibrillation ablation: A new end point for
  enhancing pulmonary vein isolation,'' {\em Heart Rhythm}, vol.~11, no.~11,
  pp.~1904 -- 1911, 2014.

\bibitem{Rodriguez_2018}
Rodríguez-Mañero {\em et~al.}, ``Validating left atrial low voltage areas
  during atrial fibrillation and atrial flutter using multielectrode automated
  electroanatomic mapping,'' {\em JACC: Clinical Electrophysiology}, vol.~4,
  no.~12, pp.~1541 -- 1552, 2018.

\bibitem{Liang_2017}
J.~J. Liang {\em et~al.}, ``Comparison of left atrial bipolar voltage and scar
  using multielectrode fast automated mapping versus point-by-point contact
  electroanatomic mapping in patients with atrial fibrillation undergoing
  repeat ablation,'' {\em Journal of Cardiovascular Electrophysiology},
  vol.~28, no.~3, pp.~280--288, 2017.

\bibitem{Magtibay_2017}
K.~Magtibay {\em et~al.}, ``Resolving bipolar electrogram voltages during
  atrial fibrillation using omnipolar mapping,'' {\em Circulation.}, vol.~10,
  no.~9, 2017-09.

\bibitem{Rasmussen_2006}
C.~E. Rasmussen and C.~K.~I. Williams, {\em Gaussian processes for machine
  learning}.
\newblock Adaptive computation and machine learning, Cambridge, Mass: MIT
  Press, 2006.

\end{thebibliography}

\end{document}